\documentclass[preprint,showpacs,preprintnumbers,amsmath,amssymb,superscriptaddress]{revtex4}

\usepackage{graphicx,subfigure}% Include figure files
\usepackage{dcolumn}% Align table columns on decimal point
\usepackage{bm}% bold math

\begin{document}
\title{Quantum critical points and phase separation instabilities in Hubbard nanoclusters}

\author{A. N. Kocharian}
\affiliation{Department of Physics, California State University, Los Angeles, CA 90032, USA}
\author{Kun Fang}
\affiliation{Department of Physics, University of Connecticut, Storrs, CT 06269, USA}
\author{G. W. Fernando}
\affiliation{Department of Physics, University of Connecticut, Storrs, CT 06269, USA}

\begin{abstract}
Spontaneous phase separation instabilities with the formation of various types of charge and spin pairing (pseudo)gaps in $U>0$ Hubbard model including the {\it next nearest neighbor coupling} are calculated with the emphasis on the two-dimensional (square) lattices generated by $8$- and $10$-site Betts unit cells. The exact theory yields insights into the nature of quantum critical points, continuous transitions, dramatic phase separation instabilities and electron condensation in spatially inhomogeneous systems. The picture of coupled anti-parallel (singlet) spins and paired charged holes suggests full Bose condensation and coherent pairing in real space at zero temperature of electrons complied with the Bose-Einstein statistics. Separate pairing of charge and spin degrees at distinct condensation temperatures offers a new route to superconductivity different from the BCS scenario. The conditions for spin liquid behavior coexisting with unsaturated and saturated Nagaoka ferromagnetism due to spin-charge separation are established. The phase separation critical points and classical criticality found at zero and finite temperatures resemble a number of inhomogeneous, coherent and incoherent nanoscale phases seen near optimally doped high-$T_c$ cuprates, pnictides and CMR nanomaterials.
\end{abstract}

\pacs{65.80.+n, 73.22.-f, 71.27.+a, 71.30.+h, 75.10.Lp, 75.10.Jm}

\keywords{Quantum critical point, Coherent pairing, Phase separation, Spin-charge liquid, Spin-charge separation, Spin magnetism, Betts lattice}

\maketitle

\section{Introduction}~\label{intro}
\begin{figure}
\begin{center}
\includegraphics*[width=20pc]{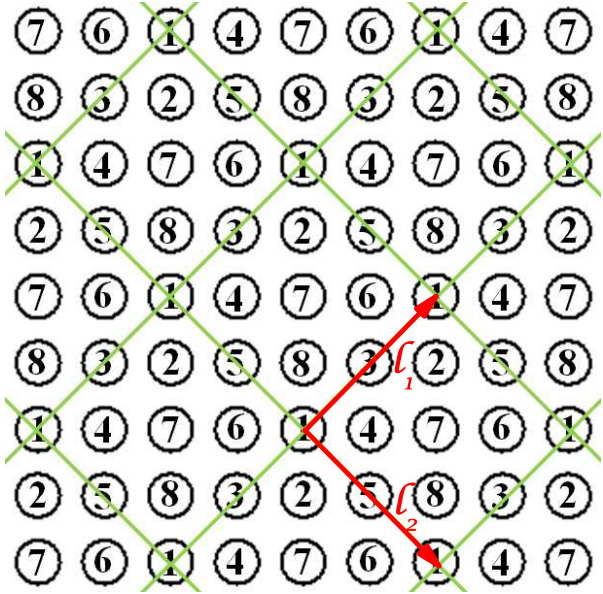}
\caption{The $8$-site finite unit cell (block) repeating periodically fill the entire ({\it infinite}) square $2d$ lattice. The cells can be defined by edge vectors $\bm l_1=(2,2)$ and $\bm l_2=(2,-2)$ (arrows in the figure) as defined in Ref.~\cite{15}.}
\label{fig1}
\end{center}
\end{figure}

A key element for understanding the complexity and perplexity in high-$T_c$ cuprates, manganites and colossal magnetoresistance (CMR) nanomaterials is the experimental observation of phase separation (PS) instabilities at the nanoscale signaled by spin-charge separation and quantum phase transitions (QPTs)~\cite{1,2,3,4,5}. A new guiding principle for the search of new materials with enhanced Tc is the proximity to quantum critical points (QCPs) for spontaneous first order QPTs attributed to intrinsic spatial inhomogeneities (see Ref.~\cite{6} and references therein). Strong quantum fluctuations dominate thermal fluctuations and affect the classical properties well above absolute zero temperature~\cite{7}. The inhomogeneous concentrated system in equilibrium can be well approximated as a quantum gas of decoupled clusters, which do not interact directly but through the grand canonical ensemble, with different electron number per cluster. Our results for possible spatial inhomogeneities are directly applicable to nanoparticles and respective bulk nanomaterials which usually contain an immense number of isolated clusters in contact with a thermal reservoir by allowing electron number per cluster to fluctuate. The finite-size optimized clusters may be one of the {\it few solid} grounds available to handle this challenging problem in a bottom-up approach~\cite{8} by defining canonical and grand canonical local gap order parameters in the absence of a long-range order, spin or charge density waves~\cite{9}. The PS instabilities and spin-charge separation effects in bipartite Hubbard clusters driven by on-site Coulomb interaction $U$ display QCPs which strongly depend on cluster topology~\cite{10,11,12,13}. In frustrated (nonbipartite) geometries spontaneous transitions depend on the sign of the coupling $t$ and can occur for all $U$ by avoiding QCPs (level crossings) at finite $U$. The existence of the {\it intrinsic QCPs} and inhomogeneities associated with the PS instabilities, are crucial ingredients of the superconducting (SC) and ferromagnetic QPTs, providing important clues for understanding the {\it incipient microscopic mechanisms} of pairing instabilities in real space due to coexisting high/low electron (hole) or high/low spin up (down) densities in high-$T_c$ superconductors (HTSCs) and colossal magnetoresistive (CMR) nanomaterials respectively. However, small systems suffer from finite-size (edge) effects, so it is unclear whether the observed instabilities can survive in the thermodynamic limit. Thus, tests on reduced boundary effects are necessary to confirm the picture of local instabilities in larger systems in the so-called ``optimized" Betts building blocks (finite square lattices)~\cite{14}. A square infinite lattice is tiled by identical square unit cells containing $L$ sites which periodically repeat in the lattice. For example, $8$-site unit Betts' cell in Fig.~\ref{fig1} is used to fill the whole lattice. The square units restore basic symmetrical properties of the infinite square lattice and periodicity of clusters partially preserves translational and rotational symmetries of the original lattice. Therefore, Betts cells are believed to be the most preferred unit blocks for relieving frustrations over other structures with the same size. All the Betts unit cells can be defined uniquely by two edge vectors~\cite{15} which represent translational vectors between the nearest two clusters. The lattice generated by Betts unit cells provides useful insights into certain physical aspects of the phase diagram in the $t-J$ model~\cite{16}. To our knowledge, an exact calculation of phase separation and pairing under doping has not been attempted in the Betts lattices applied to the Hubbard model either with nearest or next nearest neighbors. Different two-dimensional ($2d$) square structures, defined by the condition $m^2+n^2=\sqrt{\bm L}+ \sqrt{\bm L}$ with a linear size $L$ ($m$, $n$ are integers), can be used as plaquettes to extrapolate the results to the infinite square lattice. Here our primary goal is an exact study of critical instabilities in the two-dimensional $8$- and $10$-site Betts (generated) lattices.

\section{Phase separation instabilities}
The lattice in two dimensions can be tiled by periodically repeated Betts (isotropic) unit cells with reduced edge (boundary) effects.

In Fig.~\ref{fig1} an infinite square lattice has been tiled by $L$-site squares with edge vectors $\bm l_1=(2,2)$ and $\bm l_2=(2,-2)$, which represent displacements of one vertex to the equivalent vertex in the neighboring tile with which it shares an edge. Notice that, if the sites are numbered as in Fig.~\ref{fig1}, each odd site in the plaquette is surrounded by all the even sites (as nearest neighbors) and vice versa. The Betts unit cells take full advantage of the local space group symmetries of the isotropic $2d$ (square) bipartite lattice. The two-orbital Hubbard model with p bonding orbitals after elimination of the Cu sites can be reduced to a single-orbital Hubbard model with the nearest and next-nearest-neighbor (NNN) couplings by allowing holes to move within a given oxygen sublattice. Thus, we consider the minimal Hubbard model
\begin{eqnarray}
{\hat H}=-\sum\limits_{ij,\sigma}t_{ij}{\hat c}^{+}_{i\sigma}~{\hat c}_{j\sigma}+U \sum\limits_{i} {\hat n}_{i\uparrow}{\hat n}_{i\downarrow}, \nonumber
\label{eqn:h}
\end{eqnarray}
where summation goes through all lattice sites $L$ with coupling integral $t_{ij}$ equal to $t$ for the nearest and $t_nnn$ for the next nearest neighbors. The lattice frustration with $t_{nnn}\neq 0$ allows study of electron pairing in the absence of electron-hole symmetry. Below, an exact diagonalization technique is used to extract the pairing instabilities and QCPs in finite $8$- and $10$-site cluster-based Betts square lattices with periodic boundary conditions. In our previous work (Refs.~\cite{11,12,13}), we have discussed PS instabilities in selected cluster geometries. A collection of ``clusters" can be treated at a fixed average number of electrons $\langle N\rangle$ in a canonical ensemble or a fixed chemical potential $\mu$ in a grand canonical ensemble. In the canonical ensemble we define a charge gap $\Delta^c(N, T)$ at a given $U$ and temperature $T$ as $\Delta^c(N,T)=\mu_+-\mu_-$, where $\mu_+=E(N+1,T)-E(N,T)$ and $\mu=E(N,T)-E(N-1,T)$ energies are the first derivatives of the many-body (average) energy $E(N,T)$ for adding or subtracting one electron in the charge sector for the $N$-electron state. The canonical charge gap also describes electron fluctuations between different many-body cluster configurations, $d^N+d^N\rightarrow d^{N+1}+d^{N-1}$ with different electron densities, $n=N/L$. The charge fluctuations closely resemble resonant valence fluctuations~\cite{17}. Under certain conditions the charge gap can be associated with either an excitation ($\Delta^c>0$) or a binding ($\Delta^c<0$) energy.

Physically, $\Delta^c>0$ manifests a stable $d^N$ state, while $\Delta^c<0$ implies a discontinuous spontaneous PS instability, {\it i.e.}, first order transition (a negative compressibility) with the generation of $d^{N+1}$ and $d^{N-1}$ states of an inhomogeneous (mixture) of hole-rich and electron-rich phases coexisting with different $n$. Here, we focus on one hole off half filling, $\langle N\rangle =L-1$. At the same time condition $\Delta^c<0$ implies delocalization of holes (electrons) and pairing, which strongly suggests a possible local (correlation) coexistence of pairing with phase separation (segregation) rather than their competition. In the negative charge gap region ($\Delta^c<0$), the spin pseudogap is nonnegative, {\it i.e.}, $\Delta^h\geq 0$, since in the grand canonical ensemble we define it as the separation between the two consecutive peak positions of magnetic susceptibility $\chi_s=\partial S_z(\mu,h)/\partial h$ for a given $\mu$~\cite{9}. However, when $\Delta^c>0$ ($U>U_c$) we define the canonical spin gap $\Delta^s$ as an energy difference between the ground state with spin $S$ and the lowest excited state with spin $S'$ in the spin sector, $\Delta^s=E(N,S',T)-E(N,S,T)$. When the excited state has a higher (lower) total spin, the spin gap is positive (negative). In addition, we introduced the grand canonical charge pseudogap $\Delta^\mu\geq 0$ as a peak-to-peak distance between two consecutive peak positions of charge susceptibility~\cite{8,9}, $\chi_c=\partial N(\mu)/\partial \mu$. A key element for the understanding of various electron instabilities is the exact relationship between the charge gap and pseudogap $\Delta^c$, $\Delta^\mu$, and their corresponding spin counterparts $\Delta^h$, $\Delta^s$. The vanishing gaps and pseudogaps (nodes) are also crucial. As temperature approaches zero ($T\rightarrow 0$), the possible sign change in canonical gaps signifies the existence of QCPs related to first order (dramatic) changes, while zeros in the grand canonical pseudogaps, $\Delta^\mu$, $\Delta^h$, describe QCPs for continuous QPTs into quantum liquids. At finite temperatures, the vanishing gaps and pseudogaps define the classical critical temperatures (CCT) for pair condensations $T^P_c$, $T^P_s$ and crossovers $T_c$, $T^*$ into classical charge and spin liquids respectively.

\begin{figure}
\begin{center}
\includegraphics*[width=20pc]{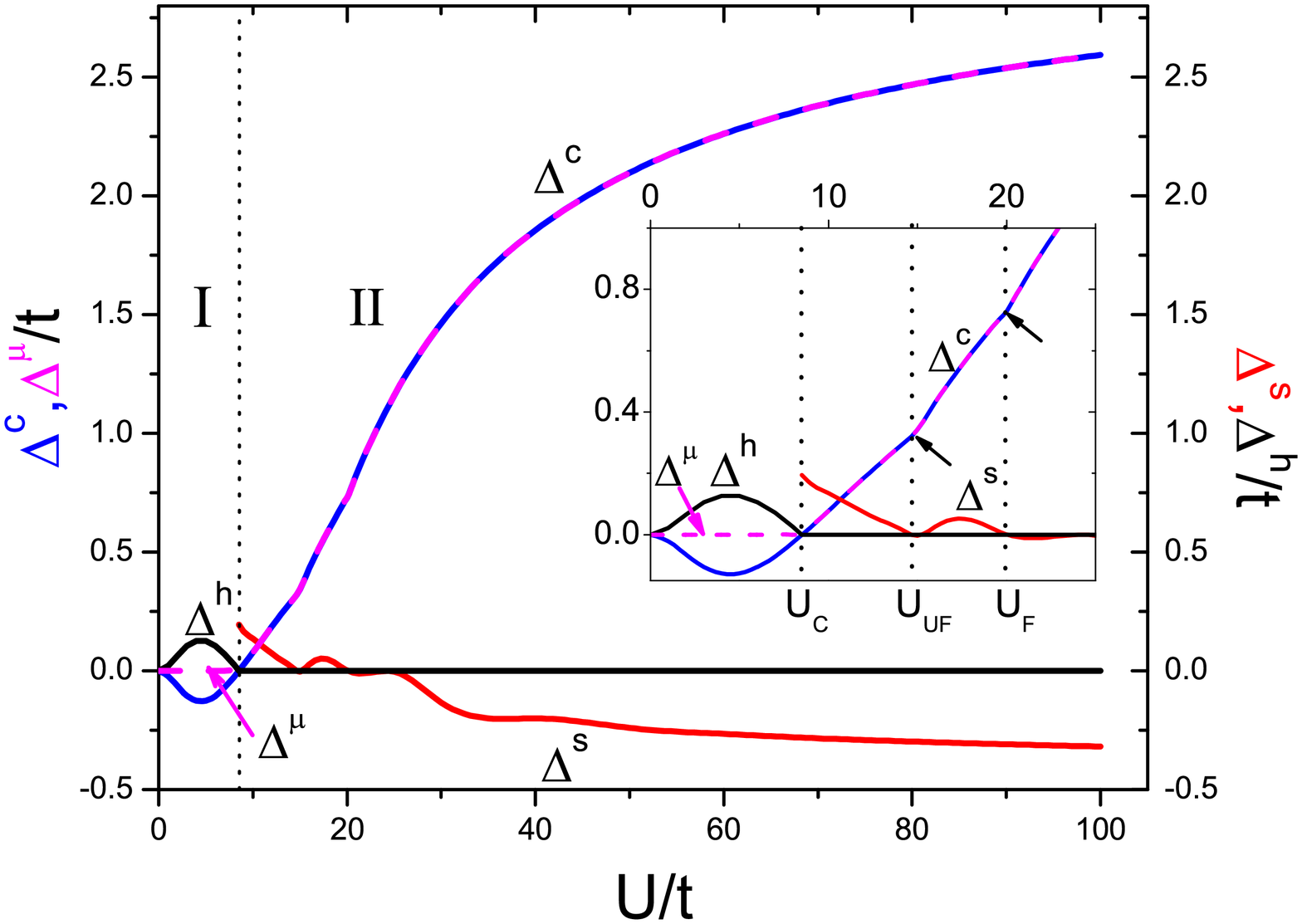}
\caption{The charge $\Delta^c$ and spin $\Delta^s$ excitation gaps in the 8-site Betts lattice for $\langle N\rangle=7$ as a function of $U$ at $t_{nnn}=0$ and $T\rightarrow 0$. The charge and spin gap nodes, $\Delta^{c,s}(U)=0$, define a number of QCPs, $U_c$, $U_{UF}$, and $U_F$ for various PS instabilities and complex phases. Phase I, $0\leq U\leq U_c$: a ground state describes coherent pairing ($\Delta^h=-\Delta^c$) of electrons (holes) with anti-parallel spins and gapless charge excitations, $\Delta^\mu\rightarrow 0$. Phase II for all $U>U_c$: a stable charge ordered MH insulator, $\Delta^\mu =\Delta^c>0$, coexists with a spin liquid, $\Delta^h\rightarrow 0$. At $U_c<U<U_{UF}$: a spin liquid phase with gapless, low spin-$3/2$ excitations. At $U_{UF}<U<U_F$: a spin liquid phase with gapless, high spin-$5/2$ excitations coexisting with unsaturated ferromagnet. At $U>U_F$: a fully saturated spin-$7/2$ ferromagnet. Inset shows an enlarged area in the vicinity of key QCPs found for elementary square geometry in Ref.~\cite{11}.}
\label{fig2}
\end{center}
\end{figure}

These pairing scenarios combined together give rise to the fascinating physics of Bose-Einstein condensation (BEC) and ferromagnetism due to competing high/low charge and high/low spin densities. The bound electron (hole) pairs with coupled anti-parallel (singlet) spins are complied with the Bose-Einstein statistics. Fig.~\ref{fig2} shows variations of gap order parameters $\Delta^c$, $\Delta^\mu$, $\Delta^h$, and $\Delta^s$ with $U$ in the 8-site Betts lattice at $T\rightarrow 0$, where we can see the number of QCPs and quantum phases. The broken line at $0\leq U \leq U_c$ for a metallic gapless charge excitation ($\Delta^\mu \rightarrow 0$) implies a quantum `cold' charge liquid behavior. A finite non-zero field is required to break the coupled anti-parallel spins, $\Delta^h>0$, which provides rigidity and stability for bound hole pairs ($\Delta^c<0$). The equal (pseudo)gap amplitudes of paired charge liquid and coupled anti-parallel spins, $\Delta^h=|\Delta^c|$, called coherent pairing in Ref.~\cite{11}, are somewhat similar to conventional BCS-type pairing with a unique quasiparticle gap.

This picture of coherent pairing of charge and spin entities in real space in Phase I (Fig.~\ref{fig2}) suggests a full BEC and possible superconductivity of electrons. In contrast, holes in Phase II, localized on separate clusters due to electron-hole pairing, $\Delta^c=\Delta^\mu>0$, at $U\geq U_c$ behave as a low density Mott-Hubbard (MH) excitonic insulator. The `cold' charge localization at the QCP, $U_c$, gives rise to a crystallized nonconducting (Mott) state driven by correlation $U$ (decreasing $t$ or pressure, $p$). Hence, $U_c$ signifies a QCP for a continuous metal-insulator (MI) transition from a charge liquid, $\Delta^\mu\rightarrow 0$, into a MH insulator. Notice that this QCP signals also spin-charge separation at $U>U_c$~\cite{8}. The evolution of spin gaps $\Delta^s$ and $\Delta^h$ is also shown in Fig.~\ref{fig2}. The isolated QCP, $U_c$, signifies a `cold' quantum melting of anti-parallel spin-insulator, $\Delta^h>0$, into a degenerate spin-liquid without rigidity ($\Delta^h\rightarrow 0$). The quantum delocalization of spin as U increases reveals an opposite trend to charge behavior. The nodes of the spin gap, defined by $\Delta^s(U)=0$, are the QCPs for the onset of spin PS and spin density (nanophase) inhomogeneities due to spontaneous redistribution of the spin liquid among the clusters in the negative gap region. Magnetic field fluctuations lead to segregation of clusters into regions rich in spin $\uparrow$ and spin $\downarrow$, {\it i.e.}, formation of magnetic domains. The QCPs for spontaneously broken symmetry at $U_{UF}=14.9$ and $U_F=19.9$ in Fig.~\ref{fig2} correspond to phase transitions into an unsaturated ferromagnet ($U_F$) and a saturated Nagaoka-like ferromagnet~\cite{11}. Phase at $U_c<U<U_{UF}$ in Table~\ref{tab1} is a degenerate spin-$3/2$ liquid without spin rigidity ($\Delta^h\rightarrow 0$). The sign change in $\Delta^s$ at $U_{UF}<U<U_F$ implies the presence of a QCP between $U_{UF}$ and $U_F$. The PS region $\Delta^s<0$ corresponds to unsaturated ferromagnetism, {\it i.e.}, $S=5/2$, while $\Delta^s>0$ region is a gapless $S=5/2$ spin liquid ($\Delta^h\rightarrow 0$) with a finite spin excitation gap ($\Delta^s>0$) for $S=7/2$.

\begin{figure}
\begin{center}
\includegraphics*[width=20pc]{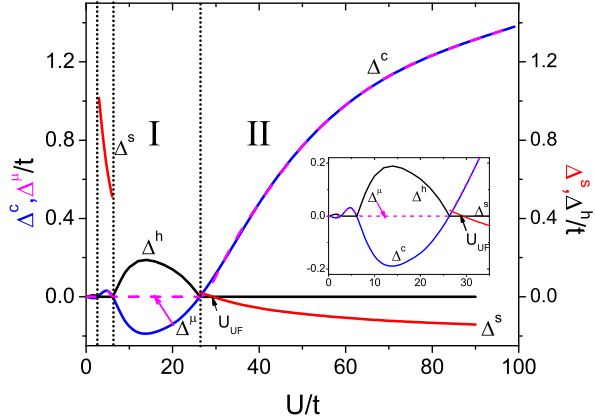}
\caption{The charge gap $\Delta^c$ as function of $U$ in the 10-site Betts lattice for $\langle N\rangle =9$ at $t_{nnn}=0$ and $T\rightarrow 0$. In contrast to the $8$-site lattice, optimized pairing is shifted to larger $U$ values. At large $U$, there is a phase transition from a MH insulator into an unsaturated ferromagnetic ground state with QCP, $U_{UF}=28.6$. Inset displays the details in a moderate $U$ region, where equal (pseudo)gaps, $\Delta^h=-\Delta^c$, display coherent pairing.}
\label{fig3}
\end{center}
\end{figure}

\begin{table}
\centering
\caption{The energy levels in the $8$-site cluster for various $S$ at $N=7$. At $Uc<U<U_{UF}$, a spin-$3/2$ liquid ($\Delta^h\rightarrow 0$) has a spin-$5/2$ excitation gap, $\Delta^s>0$. In $U_{UF}<U<U_F$ range, the state with $\Delta^s<0$ is an unsaturated $S=5/2$ ferromagnet, while next region with $\Delta^s>0$ for increasing $S=7/2$ is a gapless spin-$5/2$ liquid. At $U>U_F$, a saturated ferromagnet ($\Delta^s<0$) behaves as a spin-$7/2$ liquid.} 
\begin{tabular}{c  c   c   c   c}
\hline\hline
$U$ & $U_c<U<U_{UF}$ & \multicolumn{2}{c}{$U_{UF}<U<U_F$} & $U>UF$ \\ [0.5ex]
\hline
$\Delta^s$ & $\Delta^s>0$ & $\Delta^s<0$ & $\Delta^s>0$ &  $\Delta^s<0$ \\
\hline
\raisebox{-2.5ex}{Energy} & \line(1,0){30}\ $S=7/2$ \ \ & \line(1,0){30}\ $S=7/2$ \ \ & \line(1,0){30}\ $S=3/2$ \ \ & \line(1,0){30}\ $S=3/2$ \\
\raisebox{-1.5ex}{Levels} & \line(1,0){30}\ $S=5/2$ \ \ & \line(1,0){30}\ $S=3/2$ \ \ & \line(1,0){30}\ $S=7/2$ \ \ & \line(1,0){30}\ $S=5/2$ \\
                          & \line(1,0){30}\ $S=3/2$ \ \ & \line(1,0){30}\ $S=5/2$ \ \ & \line(1,0){30}\ $S=5/2$ \ \ & \line(1,0){30}\ $S=7/2$ \ \ \\[1ex]
\hline
\end{tabular}
\label{tab1}
\end{table}

\begin{figure}
\begin{center}
\includegraphics*[width=20pc]{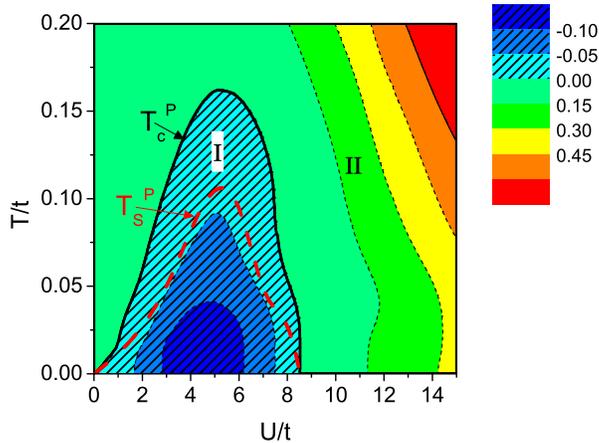}
\caption{A contour plot of the canonical charge gap as a function of $U$ and $T$ in the $8$-site Betts lattice at average $\langle N\rangle =7$. The negative gap region shrinks and totally disappears at about $T\geq 0.16$. The solid contour denotes a second order phase transition for classical critical temperature, $T^P_c(U)$, at which the gap vanishes, $\Delta^c(U,T^P_c)=0$. Phase I with a shaded area is the metallic charge pairing phase for spontaneous first order PS instabilities and density inhomogeneities. Phase II without a shaded pattern is the stable MH insulator region. The broken line for onset of coherent pair condensation of anti-parallel spins at $T_s^P(U)$ is indicative of BEC and superconductivity. Below $T_s^P$ the paired anti-parallel spin stiffness of bound charge describes coherent pairing phase, while above $T_s^P$ preformed pairs with unpaired spins is characteristics of incoherent phase~\cite{11,19}.}
\label{fig4}
\end{center}
\end{figure}

\begin{figure}
\begin{center}
\includegraphics*[width=20pc]{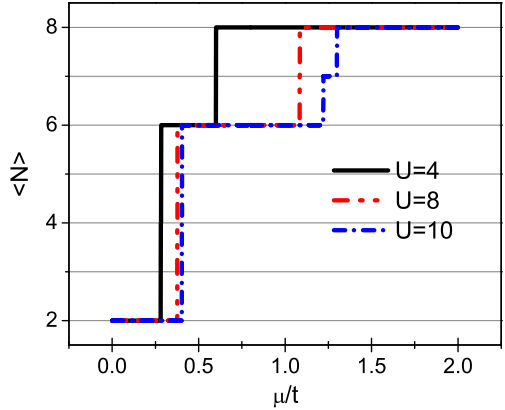}
\caption{The density $\langle N\rangle$ versus m in the $8$-site Betts lattice at three $U$ values near $U_c$ in the grand canonical ensemble at $T\rightarrow 0$. The MH plateau near $\langle N\rangle =7$ at $U=10$ vanishes with abrupt change at $U=4$ and $8$ QCPs, $\mu_P(U)$. In the canonical ensemble at $\langle N\rangle =7$ the change from $N=8$ to $N=6$ with unstable $N=7$ clusters suggests a negative compressibility for the first order PS and inhomogeneities in a spinodal region~\cite{12}.}
\label{fig5}
\end{center}
\end{figure}

\begin{figure}
\begin{center}
\includegraphics*[width=20pc]{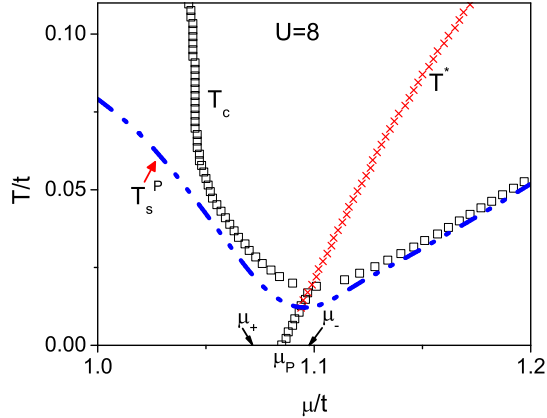}
\caption{Charge and zero field magnetic susceptibility peaks $\chi_c$ and grand canonical $\chi_s$, denoted by $T_c$ and $T^*$ in the $8$-site Betts lattice at $U=8$. The zero-field spin susceptibility peak $T^*$ terminates at $T^P_s$, while charge peak $T_c(\mu)$ for closed pseudogap, $\Delta^\mu\rightarrow 0$, terminates at the QCP, $\mu_P$. The spin peaks separate the low ($S=3/2$) and high spin ($S=5/2$) phases above and below $T^*$ respectively. With the onset of strong spin (singlet) pairing below $T^P_s$ spin collective excitations become gapped. The charge and spin peaks $T_c$ and $T^*$ emerge and reconcile for the Fermi liquid behavior at the close vicinity of $T^P_s$. The increase of the distance between $T^*$ and $T_c$, peaks above $T^P_s$ implies a spin-charge separation~\cite{9}. The charge pseudogap region due to the splitting of $T_c$ into two branches above $T^P_s$ manifests electron hole (pairing) order broken by temperature. As in~\cite{16}, the formation of incoherent charge pairs at $T^P_c\leq 0.06$ results in a charge pseudogap area above $T^P_s$.}
\label{fig6}
\end{center}
\end{figure}

Phase $U\geq U_F$: the region with $\Delta^s<0$ describes a spin liquid ($\Delta^h\rightarrow 0$) of fully polarized $S_{max}=7/2$ Nagaoka-like ferromagnet coexisting with the MH-like insulator, $\Delta^c>0$. In Fig.~\ref{fig3} the exact gaps in the $10$-site Betts lattice reproduce the key features of spontaneous PS instabilities. The negative charge gap region defines a first order PS, while the positive gap region is a signature for continuous, second order transitions. The oscillations in the Betts $10$-site lattice at small and moderate $U$ are due to reduced square symmetry similar to the anisotropic $2\time 4$ Hubbard clusters~\cite{10,11,12}. In contrast, the $8$-site Betts lattice and the $2\times 2$ Hubbard cluster both have higher rotational and reflection symmetries. Figs.~\ref{fig2} and~\ref{fig3} are consistent with our phase diagrams in $2\times 2$ and $2\times 4$ geometries~\cite{10,11}. Next we show that the effect of QCPs is also tangible over a wide range of temperatures. Contour isolines along which the charge gap attains a constant value, are illustrated in Fig.~\ref{fig4} for $t_{nnn}=0$. The contour map defines the charge gap isolines as a function of $T$ and $U$. As temperature increases charge gap changes sign and favors electron-hole pairing above the critical temperature $T^P_c$ defined by $\Delta^c(U,T^P_c)=0$. The vanishing gap describes a smooth (second order) transition for the onset of charge pairing below $T^P_c$~\cite{2,12,13}. Phase at $T>T^P_c$ is the MH insulator ($\Delta^c>0$), while at low temperature charge paired ($\Delta^c<0$) phase is a metallic charge liquid, with $\Delta^\mu=0$ (Fig.~\ref{fig6}).

The continuous correlation driven MI transition, broken by electron hole pairing at $T\geq T^P_c$, is also somewhat different from the conventional $T-p$ phase boundary between the ordered (quantum) phase at low and disordered (normal) phase at high temperatures, described by an order parameter, which is non-zero in the ordered phase and zero in the disordered phase. The charge gap shows pairing below $T^P_c$ and proximity to electron-hole pairing $\Delta^c>0$and MH pseudogap behavior ($\Delta^\mu>0$) above $T^P_c$. However, unlike in the BCS theory, as $T$ increases $\Delta^c$ differs from spin gap $\Delta^h$, consistent with the existence in Fig.~\ref{fig4} of two energy (gap order parameters or) scales and two distinct consecutive condensation temperatures for paired charges, $T^P_c$, and bound anti-parallel spins, $T^P_s$~\cite{4,11}. The spin rigidity for bound anti-parallel spins ($\Delta^h>0$) below $T^P_s$ manifests the onset of coherent pairing and possible superconductivity~\cite{2,3,11}. In Fig.~\ref{fig4} the incoherent pairing of bound charge $\Delta^c(T)\neq 0$ without rigidity $\Delta^h(T)=0$ exists at $T^P_s\leq T\leq T^P_c$. The grand canonical result for the (dramatic) abrupt doping dependent behavior of $\langle N\rangle$ at $0<U<U_c$ in Fig.~\ref{fig5} provides strong evidence for the existence of a QCP at equilibrium in $\mu$ space associated with a spontaneous first order PS. The first order transitions at $U=4$ and $8$ at $T\rightarrow 0$ display the QCP, $\mu_P$, at optimal doping. The visible plateau near $\langle N\rangle =7$ in Fig.~\ref{fig5} for $\Delta^c>0$ at $U=10$ describes the rigidity of the MH insulating charge gap~\cite{11}.

The canonical negative gap at $T=0$ displays also a PS spinodal [$\mu_+,\mu_-$] region, $\mu_\pm=\mu_P\pm\Delta^c/2$ for coexisting inhomogeneous N?6 and 8 clusters with electron redistribution~\cite{2,3,11}. The effect of quantum criticality can be felt also at low T without even ever reaching the ground state. Fig.~\ref{fig6} shows the susceptibility peak crossover $T_c(\mu)$ and $T^*(\mu)$ temperatures versus $\mu$ for gapless charge and spin excitations in the grand canonical ensemble. The peak-to-peak separation for the two splitting branches of $T_c$ beneath $T^*$ in the underdoped regime at $\mu\geq\mu_P$ defines the so-called pseudogap above $T^P_s$. The distinct pseudogap phase, fundamentally different from pairing, is a precursor to the change of the charge gap sign at low temperature, $T\leq T^P_c$~\cite{18}. We find that the pseudogap predominantly originates from $\langle N\rangle =7$ (plateau) phase at $U_c<U<U_{UF}$, which is stabilized as $T$, $\mu$ or $U$ increases or $p$ decreases~\cite{19}. In many real materials, the contribution of the $t_{nnn}$ term (frustration) can be important for breaking the particle-hole symmetry in hole/electron doped systems.

\begin{figure}
\begin{center}
\includegraphics*[width=20pc]{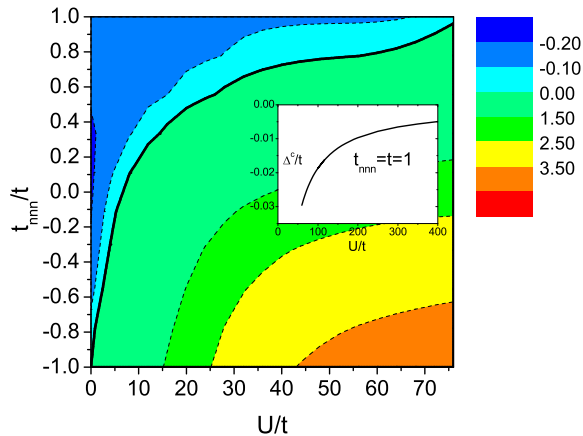}
\caption{The contour plots of charge gap $\Delta^c$ as a function of $U$ and $t_{nnn}$ in the 8-site Betts lattice. As $t_{nnn}>0$ increases, the crossover point shifts to a larger $U$ value (the inset displays the charge gap at $t_{nnn}=t=1$ in the enlarged area of $U$). As $t_{nnn}$ approaches $t$ or $-t$, PS instability region increases or shrinks respectively, by avoiding for $t_{nnn}=-1$ of level crossings (QCPs) at finite $U$.}
\label{fig7}
\end{center}
\end{figure}

Thus, it is necessary to consider the more realistic case, $t_{nnn}\neq 0$. The charge gap contour plots for different $t_{nnn}$ and $U$ values at $T=0$ are shown in Fig.~\ref{fig7} by tuning $U$ and the next nearest coupling, $t_{nnn}$. The condition $\Delta^c(U, t_{nnn})=0$ defines the boundary $t^c_{nnn}(U)$ between positive and negative gaps, which separates smooth transitions from first order phase transitions. The $t_{nnn}<0$ has a detrimental effect, while $t_{nnn}>0$ enhances pairing. The PS region shown in inset in Fig.~\ref{fig7} at $t_{nnn}=t$ manifests a coherent pairing instability for all $U$. As for the tetrahedron~\cite{13}, the vanishing PS region at $t_{nnn}=-t$, is signaled by stable $d^7$ MH configurations with $S_{max}=7/2$ ferromagnetism.

\section{Summary}
The Betts lattices, generated by finite clusters, provide strong support for PS instabilities found in generic $2\times 2$ and $2\times 4$ clusters~\cite{9,10}. Many complex critical phenomena, apparent in approximate treatments of ``concentrated" inhomogeneous systems without long-range order are naturally reproduced in the grand canonical and canonical ensembles of independent clusters, which do interact thermally. The exact calculations in optimized clusters are a powerful tool for unveiling hidden generic QCPs of phase separation, `hot' and `cold' localization and melting of charge and spin entities. The theory yields strong evidence for the existence of a QCP at optimal doping, $\mu_P$, responsible for a first order PS transition from a SC metal into a MH insulator hidden beneath the SC dome in the HTSCs~\cite{2,3,19}. The negative canonical gaps are sufficient for spontaneous PS and accompanied inhomogeneous redistribution of the electron charge or spin at energy level crossings. Small variations in model parameters or intrinsic symmetries of the unit block can lead to dramatic qualitative changes in the vicinity of QCPs, which control the physics of phase separation over a significant portion of the phase diagram. The key intrinsic QCPs, extracted from exact calculations, decipher the mystery of quantum and classical critical behavior seen in transition metal oxides, cuprates, pnictides and CMRs nanomaterials under variation of external parameters and provide a basic scenario for Nagaoka-type ferromagnetism and correlation of electron pairing with phase separation~\cite{20,21,22}. Building blocks spontaneously organized into ordered structures suggest a bottom-up key paradigm to control QCPs and fabricate of novel assembled nanostructures with new superconducting and magnetic properties~\cite{8,23,24}.

\section*{Acknowledgments}
The authors  acknowledge the computing facilities provided by the Center for Functional Nanomaterials, Brookhaven National Laboratory, which is supported by the U.S. Department of Energy, Office of Basic Energy Sciences, under Contract No.DE-AC02-98CH10886. This work was performed also, in part, at the Center for Integrated Nanotechnologies, a U.S. Department of Energy, Office of Basic Energy Sciences user facility at Los Alamos National Laboratory (Contract DE-AC52-06NA25396) and Sandia National Laboratories (Contract DE-AC04-94AL85000).


\begin{thebibliography}{11}
\bibitem{1} P.W. Anderson, Science {\bf 288} (2000) 480. 
\bibitem{2} Y. Kohsaka, {\it et al.}, Science {\bf 315} (2007) 1380. 
\bibitem{3} J. Lee, {\it et al.}, Science {\bf 325} (2009) 1099. 
\bibitem{4} S. Sachdev, Science {\bf 288} (2000) 475. 
\bibitem{5} D.M. Broun, Nature Physics {\bf 4} (2008) 170. 
\bibitem{6} A.N. Kocharian, {\it et al.}, Scanning Probe Microscopy in Nanoscience and Nanotechnology, Springer, 2010, pp. 507-570. 
\bibitem{7} Z. Fisk, Science {\bf 325} (2009) 1348. 
\bibitem{8} A.N. Kocharian, {\it et al.}, Physical Review B {\bf 74} (2006) 024511; \\A.N. Kocharian, {\it et al.}, Journal of Magnetism and Magnetic Materials {\bf 300} (2006) e585.
\bibitem{9} S. Sachdev, Physica Status Solidi B {\bf 247} (2010) 537. 
\bibitem{10} A.N. Kocharian, {\it et al.}, Physics Letters A {\bf 373} (2009) 1074; \\A.N. Kocharian, {\it et al.}, Physics Letters A {\bf 364} (2007) 57.
\bibitem{11} A.N. Kocharian, {\it et al.}, Physical Review B {\bf 78} (2008) 075431. 
\bibitem{12} G.W. Fernando, {\it et al.}, Physical Review B {\bf 75} (2007) 085109; \\K.	Palandage, {\it et al.}, Journal of Computer-Aided Materials Design {\bf 14} (2007) 103. 
\bibitem{13} G.W. Fernando, {\it et al.}, Physical Review B {\bf 80} (2009) 014525. 
\bibitem{14} J. Oitmaa, {\it et al.}, Canadian Journal of Physics {\bf 56} (1978) 897. 
\bibitem{15} D.D. Betts, {\it et al.}, Canadian Journal of Physics {\bf 74} (1996) 54; \\D.D. Betts, {\it et al.}, Canadian Journal of Physics {\bf 77} (1999) 353. 
\bibitem{16} E. Kaxiras, {\it et al.}, Physical Review B {\bf 37} (1988) 656. 
\bibitem{17} A.N. Kocharian, {\it et al.}, Soviet Physics JETP {\bf 44} (1976) 404. 
\bibitem{18} D. van der Marel, Nature Physics {\bf 7} (2011) 10. 
\bibitem{19} S.E. Sebastian, {\it et al.}, Proceedings of the National Academy of Sciences of the United States of America {\bf 107} (2010) 6175. 
\bibitem{20} Y.J. Uemura, {\it et al.}, Nature Physics {\bf 3} (2007) 29. 
\bibitem{21} E. Dagotto, Science {\bf 309} (2005) 257. 
\bibitem{22} J.M. Tranquada, {\it et al.}, Physical Review Letters {\bf 78} (1997) 338. 
\bibitem{23} K. Fang, {\it et al.}, Physics Letters A {\bf 376} (2012) 538. 
\bibitem{24} J.I. Sohn, {\it et al.}, Nano Letters {\bf 9} (2009) 3392. 

\end{thebibliography}
\end{document}